# Optimisation of Nonlinear Spring and Damper Characteristics for Vehicle Ride and Handling Improvement

**Dinçer Özcan, Ümit Sönmez, Levent Güvenç**
Automotive Control and Mechatronics Research Center
Mechanical Engineering Department, İstanbul Technical University, İstanbul, Turkey

**Ş. Server Ersolmaz, İ. Erhan Eyol**
Ford Otosan, Kocaeli, Turkey

**ABSTRACT**

In this paper, the optimum linear/nonlinear spring and linear/nonlinear damper force versus displacement and force versus velocity characteristic functions, respectively, are determined using simple lumped parameter models of a quarter car front independent suspension and a half car rear solid axle suspension of a light commercial vehicle. The complexity of a nonlinear function optimisation problem is reduced by determining the shape a priori based on typical shapes supplied by the car manufacturer and then scaling it up or down in the optimisation process. The vehicle ride and handling responses are investigated considering models of increased complexity. The linear and nonlinear optimised spring characteristics are first obtained using lower complexity lumped parameter models. The commercial vehicle dynamics software Carmaker is then used in the optimisation as the higher complexity, more realistic model. The performance of the optimised suspension units are also verified using this more realistic Carmaker model.

**INTRODUCTION**

Vehicle suspension design and performance problems have been studied extensively using simple car models such as 2 d.o.f quarter car, 4 or 6 d.o.f half car or 7 d.o.f full car models. Usually the suspension design methodologies are based on analytical methods where a linear vehicle model is investigated by solving linear ordinary differential equations. Laplace and Fourier transforms are used as very valuable tools while investigating suspension units with linear characteristics. The performance functions represented by a transfer function in Laplace and/or Fourier domain might be considered to be related to ride comfort, tire forces and handling criteria versus road roughness input to achieve an optimum design. On the other hand, the investigation of nonlinear suspension characteristics must be based more on numerical methods rather than analytical methods due to the more complicated nature of the problem. In this investigation, both linear and nonlinear spring and damper characteristics of a light commercial vehicle are considered and used in the optimisation study.

In the following subsection, the optimisation requirements of suspension systems and the state-of-the-art of suspension research in the last decade is reviewed in detail. This review includes the well-known ride and handling trade-off optimisation as well as geometrical optimisation of light commercial vehicle suspension systems. Some heavy vehicle suspension optimisation papers are also reviewed due their conceptual contribution to the subject.

**Literature Review**

Vehicle suspensions can be regarded as interconnections of rigid bodies with kinematic joints and compliance elements such as springs, bushings, and stabilizers. Design of a suspension system requires detailed specification of the interconnection points (or so called hard points) and the characteristics of compliance elements. Tak and Chung [1] proposed a systematic approach to achieve optimum geometric design of suspension systems. During the design process, these design variables are determined to meet some prescribed performance targets expressed in terms of suspension design factors, such as toe, camber, compliance steer, etc.

Koulocheris *et al.* [2] proposed to combine deterministic and stochastic optimisation algorithms for determining optimum vehicle suspension parameters. They proved that such combination yields significantly faster and more reliable convergence to the optimum. Their method combines the advantages of both categories of deterministic and stochastic optimisation. They used a half-car model of suspension systems, subject to various road profiles considering the improvement of the passengers' ride comfort, leading to minimisation of the maximum acceleration of the sprung mass, while paying attention to the geometrical constraints of the suspension as well as the necessary traction of the vehicle.

Maximising tractive effort is essential to competitive performance in the drag racing environment. Anti-squat is a transient vehicle suspension phenomenon which can dramatically affect tractive effort available at the motorcycle drive tire. Wiers and Dhingra [3] addressed the design of a four-link rear suspension of a drag racing motorcycle to provide anti-squat. This design increases rear tire traction, thereby improving vehicle acceleration performance. For the drag racing application considered, any increase in normal forces at the tire patch helps improve race competitiveness.

Mitchell *et al*. [4] used a genetic algorithm for the optimisation of automotive suspension geometries considering the description of a suspension model and a scoring method. Their approach is to design with a unit-free measure of fitness for each test and then to combine these with a weighting function. They showed that the genetic algorithm and the scoring mechanism worked effectively and significantly faster than the more common grid optimisation technique.

Raghavan [5] presented an algorithm to determine the attachment point locations of the tie-rod of an automotive suspension, in order to achieve linear toe change characteristics with jounce and rebound of the wheel. This linear behavior is advantageous for achieving good ride and handling. Raghavan's procedure can be applied to all suspension mechanism types such as short-long-arm, MacPherson struts, five-link front and five-link rear suspensions.

The design of suspension systems generally demands a compromise solution to the conflicting requirements of handling and ride comfort. The following examples demonstrate this compromise.
- For example, for better comfort a soft suspension and for better handling a stiff suspension is needed.
- A high ground clearance on rough terrain is required, whereas a low center of gravity height is desired for swift cornering and dynamic stability at high speeds.
- It is advantageous to have low damping for low force transmission to the vehicle frame; on the contrary high damping is desired for fast decay of oscillations.

Considering the above requirements, Deo and Suh [6] proposed a novel design for a customizable automotive suspension system with independent control of stiffness, damping and ride-height. This new design enables the providing of desired performance depending on user preference, road conditions and maneuvering inputs while avoiding the performance trade-offs.

Goncalves and Ambrosio [7] proposed a methodology in order to investigate flexible multibody models for the ride and stability optimisation of vehicles. Their methodology allows the use of complex shaped deformable bodies, represented by finite elements. The ride optimisation is achieved by finding the optimum of a ride index that is the outcome of a metric that accounts for the acceleration measured in several key points of the vehicle, weighted according to their importance for occupant comfort.

Duysinx *et al*. [8] developed a mechatronic approach to model, simulate and optimize a passenger car (Audi A6) incorporating a controlled semi-active suspension. They paid particular attention to the formulation of the mechatronic model of the car including:
- A mechanical sub-model; the vehicle (chassis-suspensions-wheels),
- An electro-hydraulic sub-model: to describe the behavior of the semi-active shock absorbers,
- A control sub-model tailored to satisfy comfort and/or ride criteria.

They used and compared two different modeling and optimisation approaches. The first approach is carried out in the MATLAB-SIMULINK environment and the derivation of the equations is based on a symbolic multibody model. The optimisation procedure has also been investigated in MATLAB. The second approach relies on a multibody model based on the finite element method where the optimisation has been realized with an open source industrial optimisation tool.

Eskandari *et al*. [9] optimized the handling behavior of a mid-sized passenger car by altering its front suspension parameters using Adams/Car software. They utilized an objective function combination of eight criteria indicating handling characteristics of the car and reduced the amount of optimisation parameters, by implementing the design of experiments method capabilities. The amount of the parameters was reduced from fifteen to ten by using a sensitivity analysis.

Recently, He and McPhee reviewed [10] the state-of-the-art related to modeling approaches, considering vehicle system models, design variable and performance criteria definitions, optimisation problem formulation methods, optimisation search algorithms, sensitivity analysis, computational efficiency and other related techniques. They applied these techniques to the design synthesis of ground vehicle suspensions and proposed a methodology for automated design synthesis of ground vehicle suspensions.

Papageorgiou and Smith [11] presented the development of a simulation-based methodology for the analysis and optimal design of nonlinear passive vehicle suspensions. They constructed a nonlinear vehicle model using the Matlab/Simulink toolbox SimMechanics and considered the detailed representation of the suspension geometry and the nonlinearities of the suspension elements. Several aspects of suspension performance were considered such as ride comfort, tire grip and handling which were evaluated using a simulation run in the time-domain. Their approach is similar to ours except that we consider both nonlinear damper and nonlinear spring characteristics and validate the performance of the optimized suspension units using Carmaker software.

Li *et al*. [12] considered a five-link suspension optimization for improving the ride safety and comfort using ADAMS/Insight. They investigated the relations among multi-link suspension structural parameter, wheel location parameter, and wheel track.

Uys *et al*. [13] reported an investigation to determine the spring and damper settings that will ensure optimal ride comfort of an off-road vehicle, on different road profiles and at different speeds. Spring and damper settings in the 4S4 can be set either to the ride mode or the handling mode and therefore a compromise ride-handling suspension is avoided. They found that; optimizing for a combined driver plus rear passenger seat weighted root mean square vertical acceleration rather than using driver or passenger values only, returns the best results. Their results indicated that optimization of suspension settings using the same road and constant speed will improve ride comfort on the same road at different speeds and these settings will also improve ride comfort for other roads at the optimization speed and other speeds, although not as much as when optimization has been done for the particular road.  We take into account their statement and used one vehicle speed for optimization. They also concluded that for improved ride comfort, damping generally has to be lower than the standard (compromised) setting, the rear spring as soft as possible and the front spring ranging from as soft as possible to stiffer depending on road and speed conditions. Ride comfort is most sensitive to a change in rear spring stiffness.

The roll steer of a front McPherson suspension system is studied and the design characteristics of the mechanism are optimized by Habibi *et al*. [14] using the genetic algorithm method. The roll steer affects handling and dynamic stability of the vehicle due to variation of the angles of the wheel and the suspension links (i.e. camber, caster and toe). However these changes cause other problems. In their paper [14]; Habibi et al. used a genetic algorithm method to determine the optimum length and orientation of the mechanism's members to minimize the variations of the toe, camber and caster angles. They defined a performance index which expresses the overall variations of the main parameters in the whole range of rolling of the body.

A general formulation for multibody flexible systems, with linear elastic deformations, is considered by Goncalves and Ambrosio [15] and applied to a road vehicle where flexibility plays an important role in its ride and handling dynamic behavior. Using finite elements to describe the flexibility of the body and the modal superposition method have the advantage of greatly reducing the dimensionality of the system. The presented results in [15] showed that the use of the detailed vehicle model within the framework of ride optimisation, leads to a measurable improvement of the comfort conditions for different road profiles and driving conditions.

An optimum concept to design 'road-friendly' heavy vehicles with the recognition of pavement loads as a primary objective function of vehicle suspension design was investigated by Sun [16]. A walking-beam suspension system is used as an illustrative example of vehicle model to demonstrate the concept and process of optimisation. Dynamic response of the walking-beam suspension system was obtained by means of stochastic process theory. Using the direct update method, optimisation is carried out when tire loads are taken as the objective function of suspension design. The results showed that tires with high air pressure could lead to more damage in pavement structures, and increasing suspension damping and tire damping can reduce the tire loads and pavement damage.

**The Scope of the Investigation**

In this section, the scope of the current investigation is summarized. The use of a complex three dimensional model of the vehicle, with a detailed description of all suspension systems and road/tire interaction, is necessary to fully investigate the problem. However, such models are computationally expensive especially when used in an iterative optimisation design process. A good alternative which is used here is the optimisation of a subsystem of a complex model. The suspension subsystem is very important in terms of vehicle dynamics considering the spring and damper load deflection characteristics as the basic design variables.

The ride optimisation is achieved by finding the optimum of a ride index which results from a metric that accounts for the linear and the angular accelerations of the model's suspended mass center and properly combined in a cost function, considering their importance for the comfort of the occupant. Simulations with different road profiles are performed at constant speed. The results are presented and discussed in view of the different methods used with emphasis on models and algorithms.

Two lumped parameter models are built in MATLAB considering an independent front suspension unit (a quarter car model) and a rear axle suspension unit (a half car model) of a light commercial vehicle. The model parameters are chosen to represent the Ford Transit Connect suspension mechanisms. Vertical displacement $z_s$ and acceleration $a_s$ of the suspended mass and the tire force $F_{Tire}$ of the quarter car model are considered as the key variables in ride comfort and handling, respectively. Similarly, vertical and angular displacement $z_s$ and $\phi_s$ of the mass center and the tire forces at both of the rear wheels of the half car model are selected as the key variables for the rear suspension half car model.

The quarter car and the half car rear axle suspension models are described first with linear and then with nonlinear differential equations of motion. Various road profiles are used to determine suitable linear and/or nonlinear spring and damper characteristics. Considering a nonlinear model; a suspension optimisation unit written as an m-file in MATLAB gives more insight than using commercial vehicle simulation/analysis software. In some cases; this option (nonlinear suspension characteristics optimisation) is not readily available in commercial vehicle suspension packages.

The main objective and the contribution of this investigation is to determine the optimum nonlinear functions of the damper and the spring characteristics for the improvement of the passengers' ride comfort, and vehicle handling leading to minimization of the objective functions.

Basic shapes representing the spring and the damper characteristics (force versus deflection for the spring and force versus velocity for the damper) are used according to automotive manufacturer's specifications. Basic functional shapes in each operating mode (extension or compression regions of the spring and the damper) are predetermined and the functional fits to these shapes are obtained. These functions and their linear combinations are then scaled searching for the optimum characteristics. The emphasis of this investigation is placed on finding non-symmetric optimum nonlinear functions of the spring and the damper. Optimized functional relations are then incorporated into a model built in a commercial software to evaluate the performance of the vehicle model with the optimized suspension. Carmaker software is used to study the handling behavior of the car in standard tests (double lane change, fishhook etc.). The investigation also by looks at a scenario where ride comfort and handling are simultaneously required. The cornering behavior of a road vehicle is an important performance mode often equated with handling. In order to analyze both of ride and handling requirements, a double lane change was performed after an irregular road profile with a disturbance and then the vehicle came to a stop. The simulation results of optimized nonlinear damper and nonlinear spring functions are compared with those of the optimized linear ones in simulations.

**SUSPENSION MODELS USED**

The mathematical models of a quarter car and a half car representing one of the front quarters and the rear axle suspension units are presented in this section.

**Quarter Car Model**

The quarter car model subject to road disturbances is shown in Figure 1.

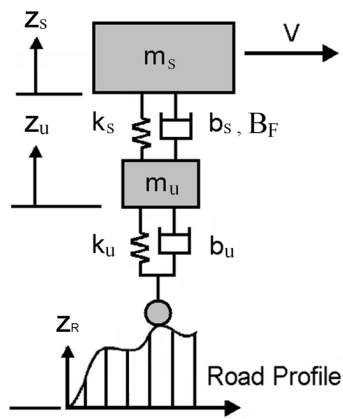

Fig.1 Quarter car model

The equation of motion considering the vertical displacement of the vehicle body with linear and nonlinear suspension characteristics may be written as

$$m_s\ddot{z}_s + b_s(\dot{z}_s - \dot{z}_u) + k_s(z_s - z_u) = 0 \qquad (1)$$

in the case of a linear suspension and as

$$m_s\ddot{z}_s + B_F(\dot{z}_s - \dot{z}_u)(\dot{z}_s - \dot{z}_u) + k_s(z_s - z_u) = 0 \qquad (2)$$

in the case of a nonlinear suspension.

Similarly the equation of motion of the vehicle wheel may be written as

$$m_u\ddot{z}_u + b_s(\dot{z}_u - \dot{z}_s) + k_s(z_u - z_s) - \ldots$$
$$b_u(\dot{z}_u - \dot{z}_r) - k_u(z_r - z_u) = 0 \qquad (3)$$

in the case of a linear suspension and as

$$m_u\ddot{z}_u + B_F(\dot{z}_s - \dot{z}_u)(\dot{z}_u - \dot{z}_s) + k_s(z_u - z_s) - \ldots$$
$$b_u(\dot{z}_u - \dot{z}_r) - k_u(z_r - z_u) = 0 \qquad (4)$$

in the case of a nonlinear suspension where $B_F(\dot{z}_s - \dot{z}_u)$ represents the nonlinear functional relation of suspension damper velocity versus force characteristic.

**Half Car Model**

The rear solid axle suspension unit of the light commercial vehicle considered here is represented with a half car model (see Fig. 2) and subjected to road disturbances coming from both sides of the track (the left and the right wheels).

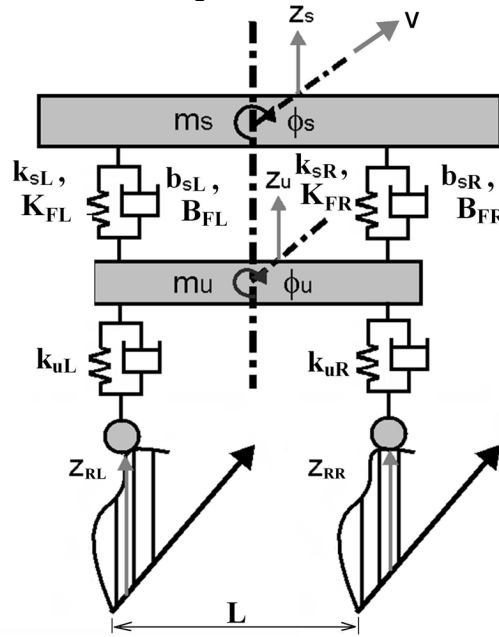

Fig. 2 Half car model

The half car model can represent the bounce ($z_s$, $z_u$) and roll motions ($\phi_s$, $\phi_u$) of the car body and solid rear axle. Therefore, it has 4 degrees-of-freedom (D.O.F).

The equations of motions of the sprung (car body) and unsprung (rear axle) masses considering the bounce and roll motions may be written as

$$m_s\ddot{z}_s + F_{sleft} + F_{sright} = 0 \qquad (5)$$

for the bounce motion of the sprung mass and

$$m_u \ddot{z}_u - F_{sleft} - F_{sright} + F_{uleft} + F_{uright} = 0 \qquad (6)$$

for the bounce motion of the rear axle.

The roll motion of the sprung mass is given by

$$I_{xx}\ddot{\phi}_s + F_{sleft}\frac{L}{2} - F_{sright}\frac{L}{2} = 0 \qquad (7)$$

and the roll motion of the rear axle (considering rotational unsprung inertia $I_{uxx}$) is represented by

$$I_{uxx}\ddot{\phi}_u - (F_{sleft} + F_{uright})\frac{L}{2} + (F_{sright} + F_{uleft})\frac{L}{2} = 0 \qquad (8)$$

where the forces $F_{sleft}$, $F_{sright}$ acting on the sprung mass are given by

$$F_{sleft} = b_{sL}\left\{(\dot{z}_s + \dot{\phi}_s\frac{L}{2}) - (\dot{z}_u + \dot{\phi}_u\frac{L}{2})\right\} + \ldots$$
$$k_{sL}\left\{(z_s + \phi_s\frac{L}{2}) - (z_u + \phi_u\frac{L}{2})\right\} \qquad (9a)$$

$$F_{sright} = b_{sR}\left\{(\dot{z}_s - \dot{\phi}_s\frac{L}{2}) - (\dot{z}_u - \dot{\phi}_u\frac{L}{2})\right\} + \ldots$$
$$k_{sR}\left\{(z_s - \phi_s\frac{L}{2}) - (z_u - \phi_u\frac{L}{2})\right\} \qquad (9b)$$

in the linear case and by

$$F_{sleft} = B_{FL}\left\{(\dot{z}_s + \dot{\phi}_s\frac{L}{2}) - (\dot{z}_u + \dot{\phi}_u\frac{L}{2})\right\} + \ldots$$
$$K_{FL}\left\{(z_s + \phi_s\frac{L}{2}) - (z_u + \phi_u\frac{L}{2})\right\} \qquad (10a)$$

$$F_{srigth} = B_{FR}\left\{(\dot{z}_s - \dot{\phi}_s\frac{L}{2}) - (\dot{z}_u - \dot{\phi}_u\frac{L}{2})\right\} + \ldots$$
$$K_{FR}\left\{(z_s - \phi_s\frac{L}{2}) - (z_u - \phi_u\frac{L}{2})\right\} \qquad (10b)$$

in the nonlinear case. Forces $F_{uleft}$ and $F_{uright}$ can be written as

$$F_{uleft} = k_{uL}(z_u + \phi_u\frac{L}{2} - z_{RL}) \qquad (11)$$

$$F_{uright} = k_{uR}(z_u - \phi_u\frac{L}{2} - z_{RR}) \qquad (12)$$

where $L$ stands for the track width, and $I_{xx}$ and $I_{uxx}$ represent the moments of inertia of the sprung mass and axle, respectively. In the simulations and optimisation process, spring and damper characteristics of the left and right sides are assumed to be identical.

## PERFORMANCE CRITERIA AVAILABLE IN THE LITERATURE

The choice of sub-objective functions and their weights in the combined (main objective) function plays a very crucial role in the optimisation process. In this section, the literature review of the vehicle suspension objective functions and the performance indices are presented in detail and our approach to the objective function formulation is presented.

In ref. [2], the non-linear stiffness and damping characteristics are optimized considering a half car model subject to different type of road irregularities. As an objective function, the maximum value of vertical acceleration of vehicle body at the passenger seats are minimized from the view point of ride comfort under the geometrical constraints of the car. The objective function is formed according to the quadratic penalty given by

$$f(x) = \max(\ddot{x}_s) + M \sum c_i^2(x) \qquad (13)$$

where $\ddot{x}_s$ is the vertical acceleration of the vehicle mass, $M$ is the penalty parameter and $c_i$ are the constraint functions for parameter vector $x$.

In reference [4], geometrical parameters of the suspension were considered when determining the fitness of a given suspension design. Since these parameters are not all at the same magnitude or even the same units; coming up with a single fitness value is difficult. The basic approach of reference [4] was to carry out the design with a unitless measure of fitness for each test and then to combine these results with a weighing function. Several functions were analyzed and compared while evaluating the speed and the accuracy of the method using the genetic optimisation algorithm. A first order normal distribution was chosen, due its convergence speed. The evaluation is calculated with the following formula,

$$score = 1 - e^{-\frac{C - bound_L}{3(C - x_i)}} \qquad (14)$$

if $x_i \leq 0$ or

$$score = 1 - e^{-\frac{bound_R - C}{3(x_i - C)}} \qquad (15)$$

which otherwise, equals *100* for the ideal score and *28.3* at the bound.

Each metric score is combined by way of a weighing function. Then, the scoring metric and total score is normalized using

$$TotalScore = \frac{\sum_i W_i score_i}{\sum_i W_i} \qquad (16)$$

The coordinates of the front and rear suspension hard points, the stiffness and damping properties of the front and rear suspension springs and damper, sprung mass, gear ratio and the inertia of steering wheel etc. were selected in ref. [17] as the design variables, considering vehicle handling. The objective evaluation index was adopted to evaluate the performance of vehicle handling. The index included: course following indices, driver burden indices, indices for the risks of roll over, index for driver's road feeling, index for lateral slip. The double lane change maneuver was selected for a virtual test and the objective evaluation index was calculated.

The weighed accelerations of the sprung mass, namely the heave ($\ddot{z}$), pitch ($\ddot{\theta}$) and roll ($\ddot{\phi}$) accelerations were considered in reference [11] for the performance measure of ride comfort. The following formula was used for the performance index

$$I_1 = \sqrt{\ddot{z}^T \ddot{z} + \ddot{\theta}^T \ddot{\theta} + \ddot{\phi}^T \ddot{\phi}} \qquad (17)$$

where the acceleration weights (see Fig. 3) were chosen according to the British Standard BS 6841 [18]. The performance measure for tire grip, considering the tire forces at the four wheel stations, was also taken into account in a time domain simulation.

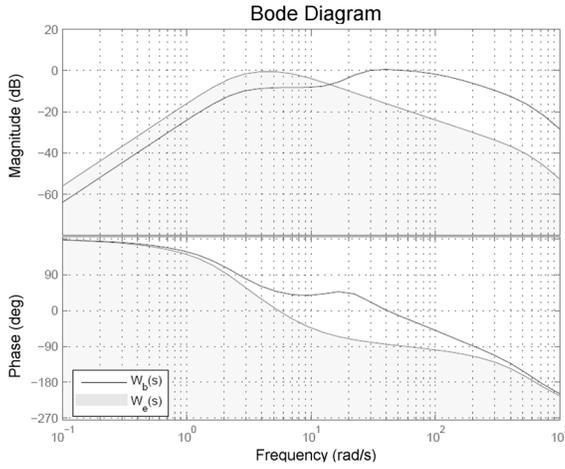

Fig.3 Frequency weights for bounce and pitch/roll discomfort as specified in British Standards [18], adapted from [11].

The objective function in reference [9] was selected representing several aspects of the handling behavior of a vehicle where the linear combination of some ride quantities with corresponding weighting factors in the following form is used.

$$F = \sum_i W_i X_i \qquad (18)$$

where $X_i$ represent; yaw velocity overshoot, yaw velocity rise time, lateral acceleration overshoot, lateral acceleration rise time, roll angle steady state response, RMS of the under-steering coefficient, RMS of the steering torque, RMS of the steering sensitivity. Selection of the weighting factors $W_i$ were made based on the importance of each quantity and are adjustable.

A global performance index was used in reference [1] as the linear combination of each individual performance index. Through kinematic analysis, toe and camber curves were obtained, and target values of the toe and camber curves were set up. The squared value of the reaction force at each tie-rod was also included in the performance index. The global performance index was determined as the weighed linear combination of the wheel angles and reaction forces at the joints.

$$I = W_1 I_{toe} + W_2 I_{camber} + \ldots + W_j I_{rollrate} + \\ \ldots + W_k I_{reactionforces} \qquad (19)$$

The performance of the active suspension system was evaluated in reference [19], covering comfort and road holding capabilities as well as the energy consumption of the system. The formulation of three different performance indices was considered; two of them are based on the RMS norm

$$|x|_{rms} = \sqrt{\frac{1}{\tau} \int_0^\tau x^2 dt} \qquad (20)$$

Comfort is strongly related to the accelerations of the vehicle body. Therefore, the performance index for comfort is formulated considering accelerations. The comfort index for a vehicle with an active suspension system is weighed in reference [19] with respect to the acceleration of the body in a conventional system and described as a ratio. A value above 1 means that the current design is inferior as compared to the passive suspension.

$$I_1 = |\ddot{z}_{active}|_{rms} / |\ddot{z}_{passive}|_{rms} \qquad (21)$$

Road holding capability is directly related to the variation in vertical tire force. A constant tire force is ideal. Additionally the index in reference [19] is weighed with respect to the passive suspension.

$$I_2 = \frac{\left|F_{T,active}\right|_{rms}}{\left|F_{T,passive}\right|_{rms}} \qquad (22)$$

where the tire force is given by

$$F_T = k_T \cdot (z_R - z_1). \qquad (23)$$

The third index $I_3$ of the objective function considers the energy used by the active system. Then, these indices are combined to form an overall objective function for the optimisation algorithm,

$$I = W_1 I_1 + W_2 I_2 + W_3 I_3 \qquad (24)$$

**PERFORMANCE INDEX USED**

In order to optimize ride characteristics, human sensitivity to vibrations needs to be considered. The motion is weighed according to the ISO 2631 standard (see ref. [20]). The different characteristics of the excitation, including magnitude, frequency, axis and duration based on the human tolerance for vibrations should be considered. As suggested by the ISO 2631 standard, the complete acceleration time histories for each of the target points are measured. Then, each Cartesian component of the acceleration histories is decomposed into a Fourier series. After that, a frequency weight given in ISO 2631 standards is multiplied by each term of the Fourier series. The single objective function value is determined as the sum of the weighted terms of the Fourier series previously obtained in the decomposition process. Frequency weights of acceleration as specified in ISO 2631-1 standards are shown Fig. 4.

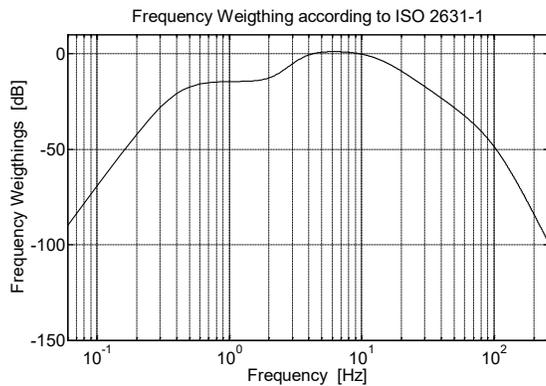

Fig.4 Frequency weights as specified in ISO 2631-1 standards.

Figure 5 shows the body acceleration response of a quarter car model to a chirp (swept sine) signal, its power spectral density (PSD) and their weighed counterparts obtained using ISO 2631 standards. The original and the weighed signals are presented in time domain comparing their magnitudes and in the frequency domain comparing their power spectral densities (see Fig. 5).

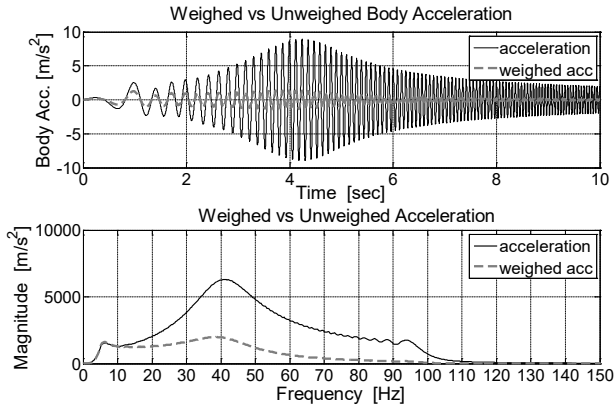

Fig. 5 Body acceleration signal is weighed according to ISO2631-1 standards

The target accelerations of the vehicle models are weighed according to the ISO 2631 standards in this investigation. There are three cases belonging to each vehicle model (total of six cases) that are investigated. The performance indices of these cases are presented in the simulation sections.

**OPTIMISATION PROCEDURE**

In this section, the optimisation procedure is explained in detail. In the current investigation, the complexity of the optimisation problem is reduced by deciding on the basic shape/behavior of the force versus displacement and force versus velocity characteristics and then scaling them up or down during optimisation. This procedure has two steps.

1. A function like a polynomial, rational or an exponential function that can fit the basic initial data of force versus displacement/velocity profile is chosen first.

2. Then, a scaling factor $C_{opt}$ is used for the function such that

$$F_{opt}(x) = C_{opt} F(x) \qquad (25)$$

Our approach includes both time and frequency domain analysis and can be summarized with the following steps.

1. First, the simulation of the quarter and the half car models subjected to a road excitation is carried out in the time domain. Note that while running the optimisation routines of the vehicle suspension models, two aspects could significantly affect results and might cause errors in the optimisation process.

   a) It is preferable to consider the steady state response of the vehicle run. Since a constant vehicle speed is assumed, it takes some time for the vehicle to reach the steady state conditions. Therefore, the beginning part of the time domain simulation containing the transient response is omitted.

   b) Attention should be paid to the static deflections (due to weights) and the initial conditions considering the static equilibrium points for the springs.

2. Then, the target point's accelerations are weighed according to ISO-2631 standards in the frequency domain, and used as part of the objective function.

3. A suitable global objective function is established according to the needs of automotive manufacturers. This is the most subjective step of the methodology. Since the choice of the objective function and weighing of the particular objectives will result in different optimum outcomes. Our choice of objective functions and performance indices for the current paper are explained in the previous section and in the following section on optimisation results.

4. As the final step, the optimisation type and algorithm are selected and the optimisation step is performed in MATLAB. The optimisation toolbox SQP algorithm with Quasi-Newton line-search is implemented. The SQP algorithm like Simplex, Complex, and Hook-Jeeves belongs to the family of local search algorithms. The local search algorithms converge to the nearest optimum, since they depend upon the starting values of the design variables. Examples in the following section illustrate the simulation result for quarter car and half car vehicle models used here. Numerical simulation results show that the SQP algorithm can efficiently and reliably find the optimum in the neighborhood of the initial point.

Finally, the optimum spring and damper characteristics obtained should be checked to see if they are manufacturable.

The nonlinear damper characteristic of the front independent suspension and rear solid axle are presented in normalized form in Fig. 6a and 6b, respectively. Since the shape of the curve is essential for manufacturing; an appropriate functional representation should be used in the optimisation process. A function with the following form is suitable for the whole range of the damper data.

$$F(v) = Ae^{-kv} + Be^{qv} \qquad (26)$$

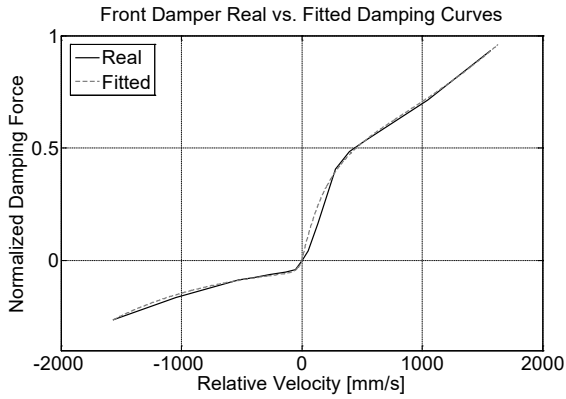

Fig. 6a Normalized damper characteristic and its functional fit (front)

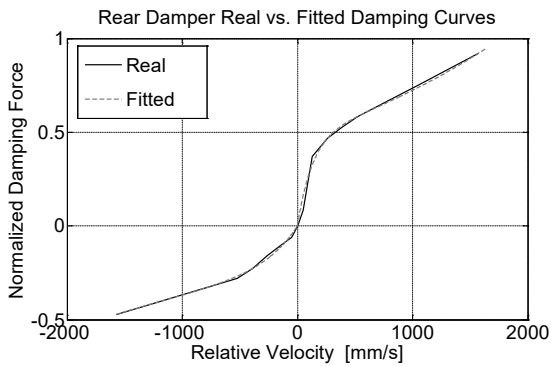

Fig. 6b Normalized damper characteristic and its functional fit (rear)

For the nonlinear modeling of the spring of the rear axle, a look-up table which represents the nonlinear characteristic of the spring is used (see Fig. 6c). The spring characteristic of the front independent suspension is linear in all optimisation processes.

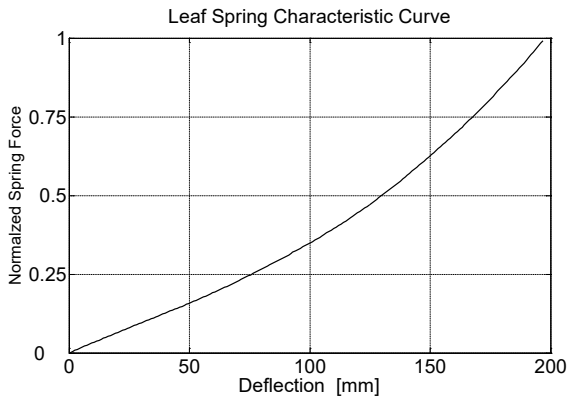

Fig. 6c Normalized spring characteristic (rear)

## OPTIMISATION RESULTS

### Quarter Car

Three different cases of optimisation runs are performed using the quarter car model in this part of the investigation including: 1) a linear suspension model subjected to a road represented by a chirp (swept sine) signal, 2) a randomly generated road and, 3) the same vehicle with a nonlinear damper and linear spring unit subjected to a randomly generated road.

Case 1: A chirp (swept sine) signal with a frequency range 0.1-20 Hz. is used as the road input. The objective function has two components which are weighed body acceleration RMS values ($I_1$) and the penalty function of the tire acceleration ($I_2$).

$$I = w_1 I_1 + w_2 I_2 \qquad (27)$$

The first term $I_1$ is the RMS of the weighed body acceleration $\ddot{Z}_s$. The performance index also contains the difference between the current tire acceleration and the desired tire acceleration ($I_2$). The desired acceleration function is the acceleration of tire mass response of a known linear suspension unit with desirable properties, obtained by the simulation process.

$$\begin{aligned} I_1 &= \ddot{Z}_s \equiv \left|\ddot{z}_s\right|_{RMS} \\ I_2 &= \ddot{z}_u - \ddot{z}_{udesired} \end{aligned} \qquad (28)$$

The total objective function changes as shown in Figure 7 during the optimisation run and an optimum point is reached after 48 iterations. Figure 8 illustrates the normalized values of the design variables, the spring stiffness and damper characteristic scaling coefficients for the quarter car model. The spring stiffness and the damper cgaracteristic scaling coefficients increase until an optimum point is reached. Larger spring and damper coefficients result in better handling performance. Figure 9 shows the $I_1$ and $I_2$ values in the optimisation process.

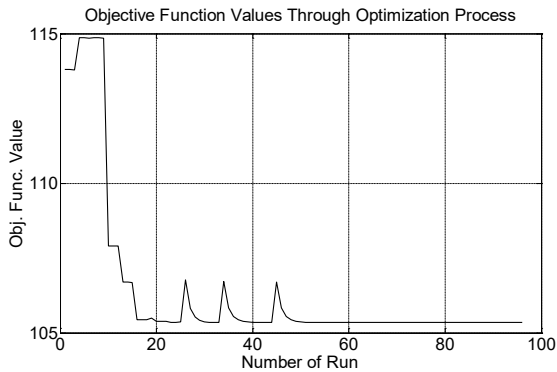

Fig. 7 Objective function's run history

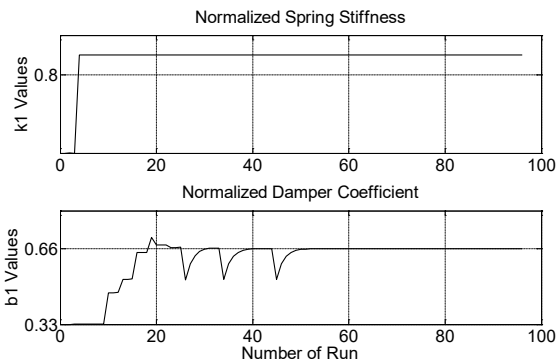

Fig.8 Optimum normalized linear spring stiffness and linear damper characteristic scaling coefficients

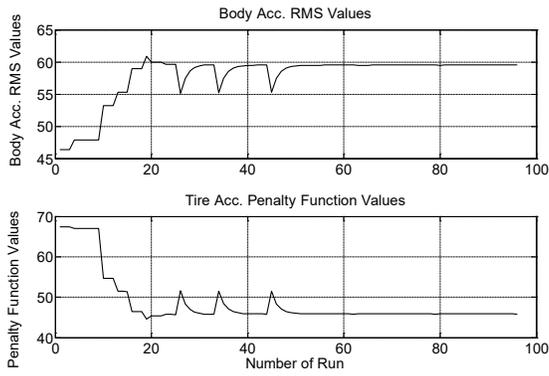

Fig. 9 Quarter car body acceleration and penalty function values for chirp road input

Bode diagrams are obtained numerically from simulation of the displacement responses of the sprung (vehicle body) and the unsprung mass (tire mass). Bode diagrams of body and tire displacements shown in Figures 10 and 11 indicate improvement in vibration magnitudes at natural damped frequencies but deterioration in between.

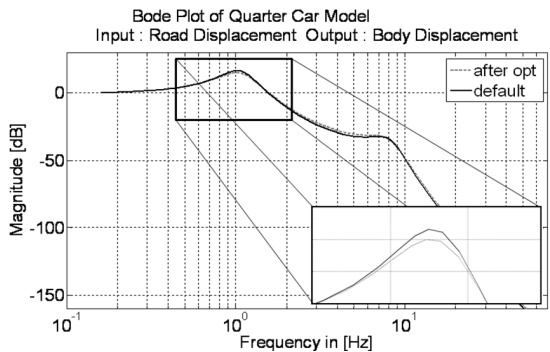

Fig. 10 Bode diagram of the car body

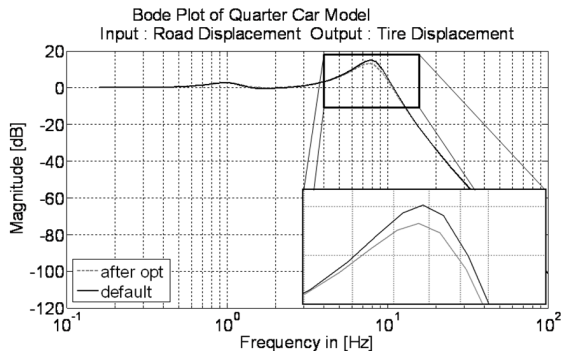

Fig. 11 Bode diagram of the tire displacement

The optimized vehicle suspension and the initial values are compared on a road represented by a chirp signal. The results displayed in Figures 12 and 13 for the body and the tire acceleration histories show the enhancement after optimisation.

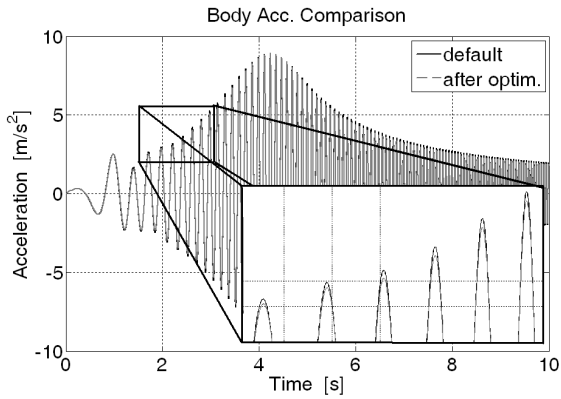

Fig.12 Body acceleration histories of the optimum suspension and the initial suspension

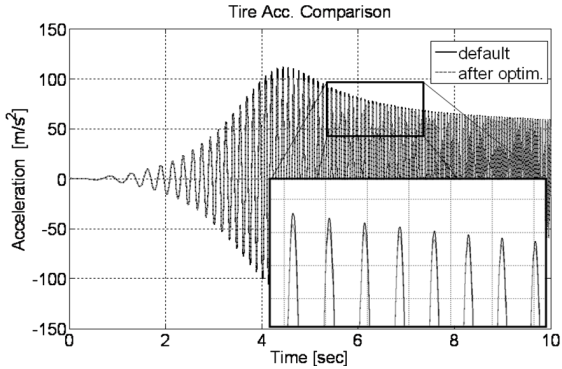

Fig.13 Tire acceleration histories of the optimum suspension and the initial suspension

Case 2: Optimisation of the quarter car with a linear suspension unit has also been studied subject to a random road profile. Figure 14 shows the road profile used. In addition, in the current case, the performance index used is the weighed linear combination of RMS values of weighed body acceleration and the change of tire forces.

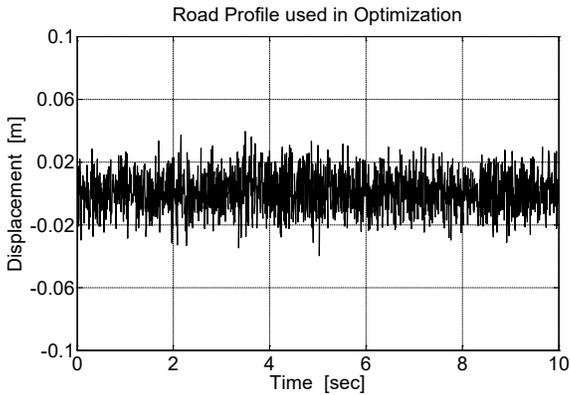

Fig. 14: Random road profile used.

The performance index

$$I = w_1 \ddot{Z}_s + w_2 \Delta F_{tire} \qquad (29)$$

where $d^2Z_s/dt^2$ is the *RMS* value of the weighed body acceleration and $\Delta F_{tire}$ is the difference between the maximum and minimum tire force picked from the tire force history. Figures 15-17 show the changes in the performance index, in the linear spring stiffness and damper characteristic scaling coefficients and in the RMS body acceleration and in the tire force change through the course of the optimisation run.

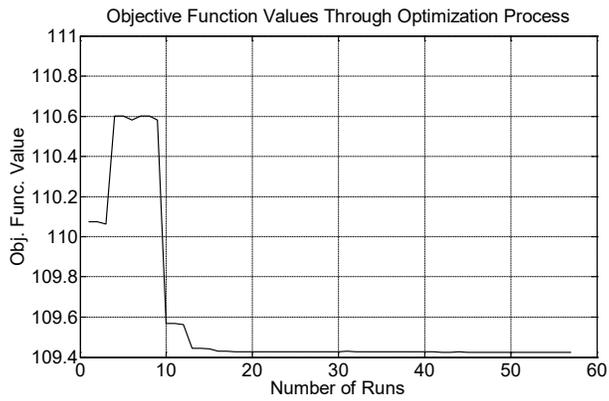
Fig.15 Objective function progress during optimization

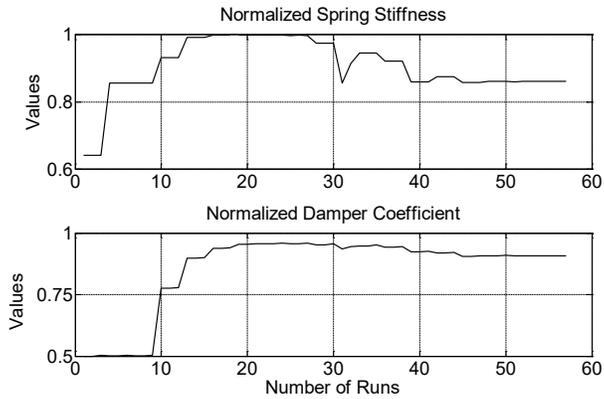
Fig.16 Linear spring and damper caharcteristic scaling coefficient progress

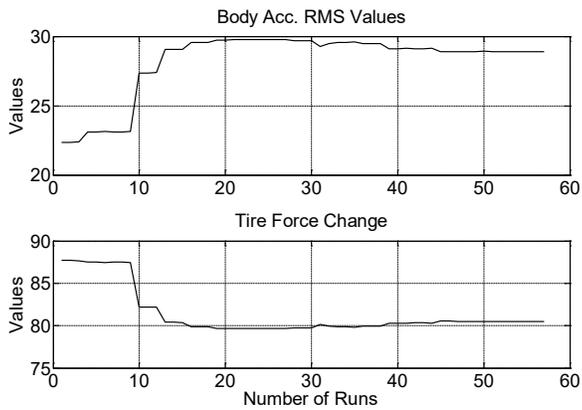
Fig.17 RMS of weighed body acceleration and tire force change

Case 3: In this case, a nonlinear damper unit is employed and the random road profile in Case 2 is used. Figure 18 displays the changes in the performance index during the course of the optimisation run. How the linear spring and nonlinear damper scaling coefficients change during the optimisation are shown in Figure 19. The two parts of the performance index are shown separately in Figure 20. Figure 21 compares the initial damper characteristic curve to the optimized one. A slight decrease in the characteristic curve is observed according to the initial curve.

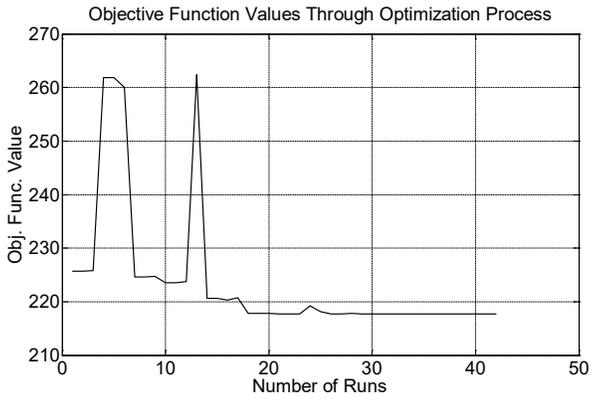
Fig.18 Total objective function

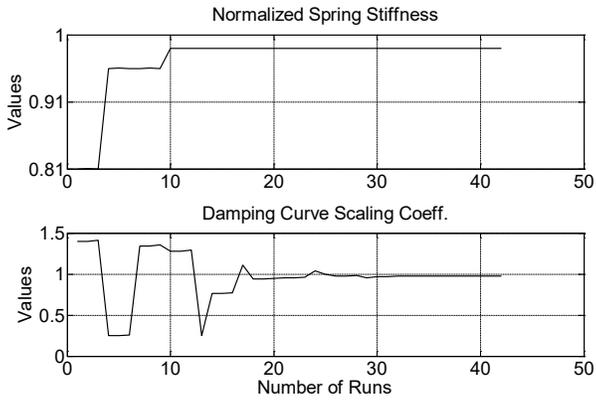
Fig.19 Linear spring stiffness coefficient and nonlinear damper characteristic scaling coefficients

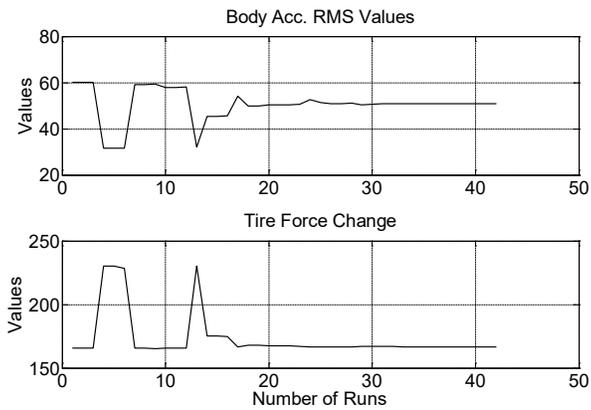
Fig. 20 Weighed body acceleration and tire force change

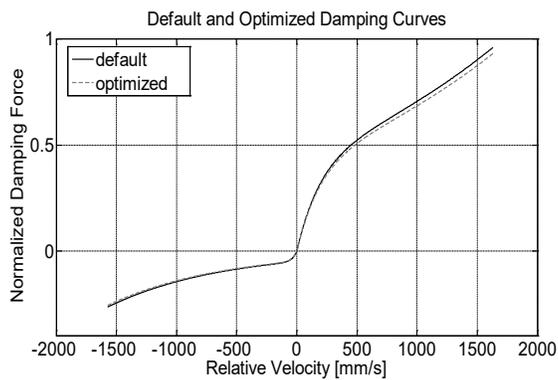
Fig. 21 Initial and optimized damper characteristic curves

**Half Car Model**

Similar to the previous subsection, three cases of the half car model embodying a rear suspension unit of the light commercial vehicle are investigated here.

Case 1: Half car with linear suspension unit optimisation results are presented in this section. The total objective function is defined as

$$I = w_1 \ddot{Z}_s + w_2 \Delta F_{tire} + w_3 \Phi_s \qquad (30)$$

where $\Phi_s$ is the RMS values of the roll angle of the body. The difference in the performance index for this case lies on the existence of the roll angle. However, for this first case, the weighing of this term $w_3$ is set to zero. The roll dynamics of the vehicle will be considered in Case 3 of this subsection.

The results for this case are presented in Figures 21-24. The change of the performance index during the course of the optimisation run is given in Figure 21 while the change of design variables are illustrated in Figures 22. and 23. Figure 24 shows the objective function values in the pre-specified region of the design variables. The minimum point of the three dimensional shape gives the optimum values of the spring stiffness and the damper coefficient which are also obtained through optimization process.

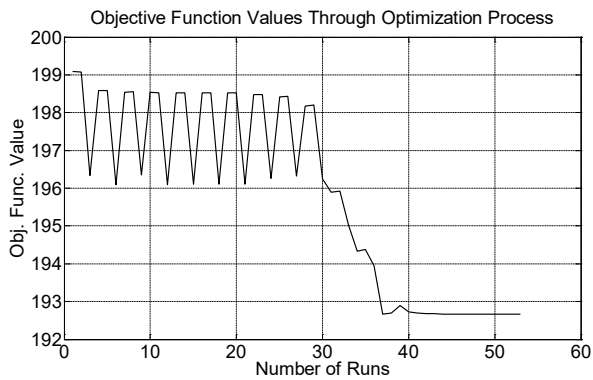

Fig. 21 Total objective progress

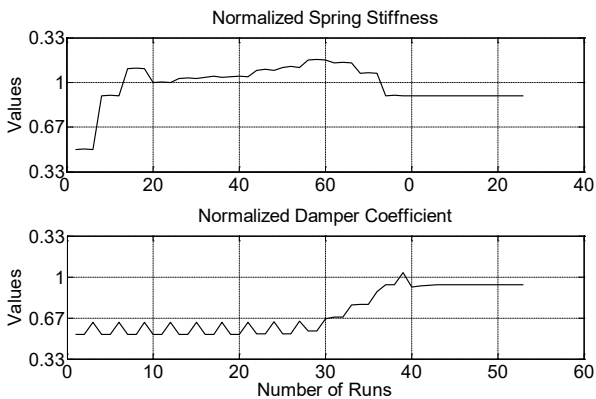

Fig. 22 Spring stiffness and damper characteristic scaling coefficient progress

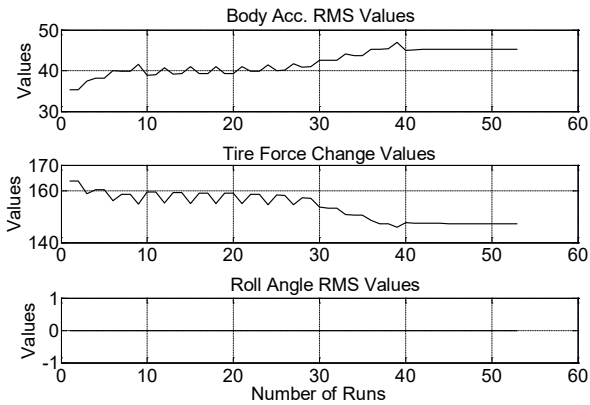

Fig.23 Objective function components' progress

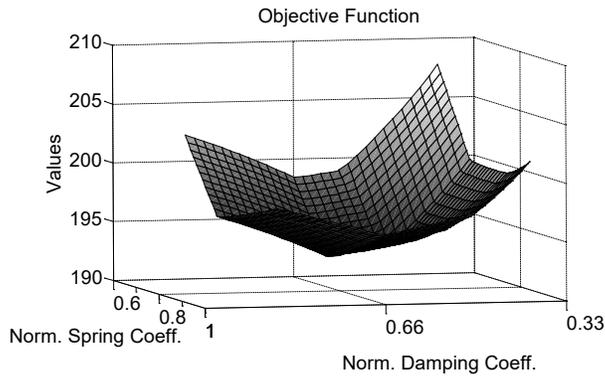

Fig. 24 Objective function values within desired range of optimisation inputs

Case 2: Nonlinear rear suspension unit incorporating nonlinear dampers and nonlinear springs whose basic characteristic curve shapes are given in the previous subsections are considered. The performance index used in the previous case is also used here. The difference between the current and the following case is in the performance index used. The optimized results of the current case are used to establish the next case's performance index.

Figure 25 displays the changes in the performance index during the course of the optimisation run. How the nonlinear spring and nonlinear damper scaling coefficients change during the optimization is shown in Figure 26. The changes of components of the performance index are shown separately in Figure 27. Figure 28 shows the objective function values in the pre-specified region of the design variables for this case. The minimum point of the three dimensional shape gives the optimum values of the spring curve scaling coefficient and the damper curve scaling coefficient which are also obtained through the optimisation run. Figures 29 and 30 show the initial and optimized damper curves and spring curves respectively. Base on these figures, it is seen that after the optimisation the damper characteristic tends to be softer while the spring characteristic tends to be stiffer.

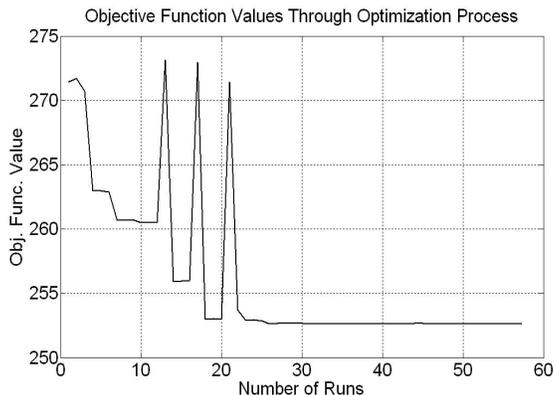

Fig.25: Objective function progress

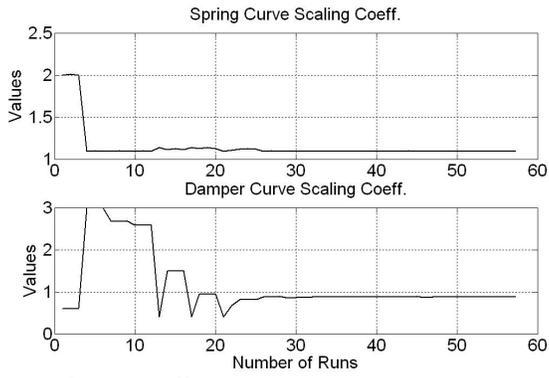
Fig. 26 Spring stiffness and damper scaling values progress

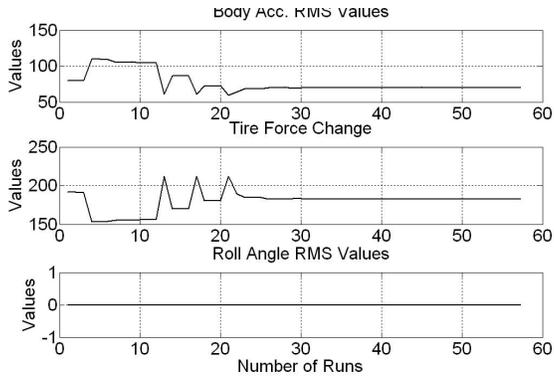
Fig.27 Components of the total objective functions

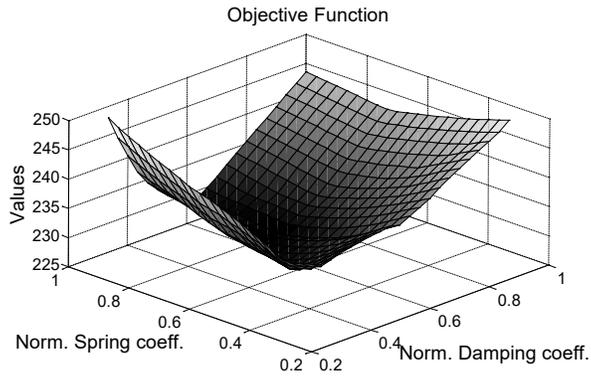
Fig. 28 Objective function values respect to scaling factors

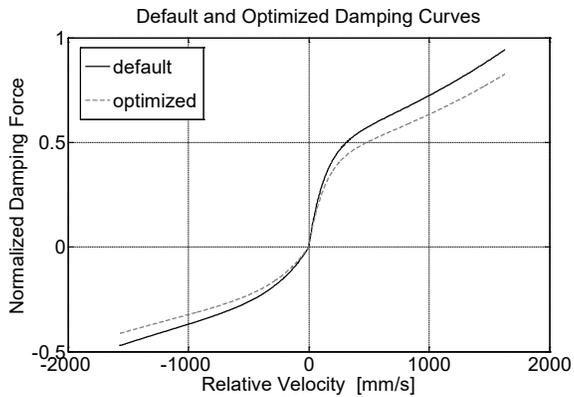
Fig.29 Optimized and default damper characteristics

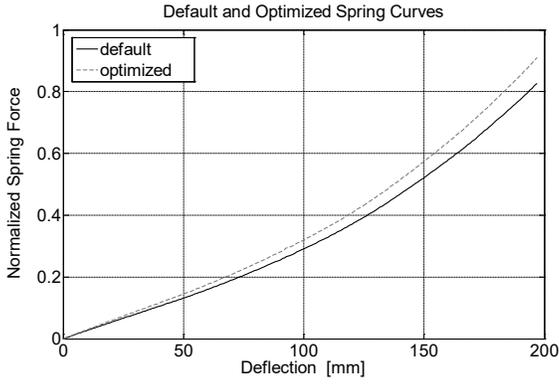

Fig. 30 Optimized and the default spring characteristics

Case 3: The only difference between this and the previous case study is the characterization of the objective function. The spring and damper characteristics similarly remain nonlinear in the current case. Unlike Case 2, the weighing of the term ($w_3$) which considers the roll dynamics is nonzero for the current case 3. Thus roll dynamics of the vehicle is taken into account in this case of the study. The time history results of the previous optimized case is used to define the objective function in case 3 as

$$I = w_1 C_{lost} + w_2 H_{lost} + w_3 \Phi_s \qquad (31)$$

where $C_{lost}$ is the comfort loss defined by the difference of the weighed *RMS* acceleration of the body in the current case and 1.1 times of that of the previous optimized case (case 2). The concept behind the new objective function $I$ is: The vehicle body acceleration history is allowed to exceed 10 % of that of the previous case. An excess of more than 10% is penalized. In other words, %10 deterioration of the comfort index is allowed during this procedure.

$$C_{lost} = \ddot{Z}_s - 1.1 \ddot{Z}_{sopt} \qquad (32)$$

where $\ddot{Z}_{sopt}$ is the optimized weighed *RMS* body acceleration of the previous case, Case 2. $\ddot{Z}_s$ is the actual weighed *RMS* body acceleration value during the optimisation run.

$H_{lost}$ corresponds to the handling loss and it is similarly defined by the difference between the current case tire force change and 1.1 times that of the previous optimized case. This can be explained as %10 deterioration of the handling index being allowed during the optimisation procedure.

$$H_{lost} = \Delta F_{tire} - 1.1 \Delta F_{tireopt} \qquad (33)$$

where $\Delta F_{tireopt}$ is the optimized tire force change of the previous case, Case 2 and $\Delta F_{tire}$ is the actual tire force change values during the optimisation run.

In order to observe and to add the influence of the roll angle in the optimisation process, a rougher road is chosen for this case.

Figures 31-34 display the results of this case. The change of the performance index values during the course of the optimisation is given in Figure 31 while the change of design variables, scaling coefficient of the spring curve and of the damper curve are illustrated in Figure 32. The initial values of the scaling factors for optimisation are taken to be identical to those obtained as the results of the previous case, Case 2. Figure 33 shows how the components of the performance index vary during the process.

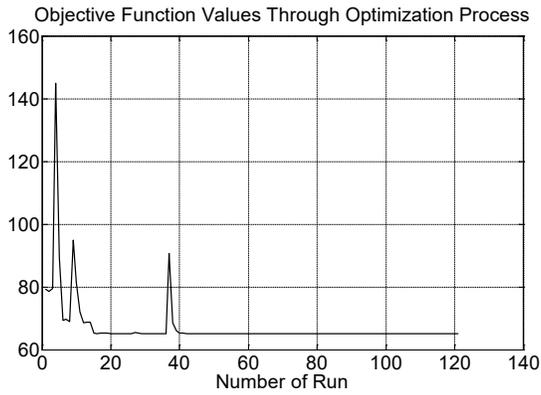
Fig. 31 Objective function progress

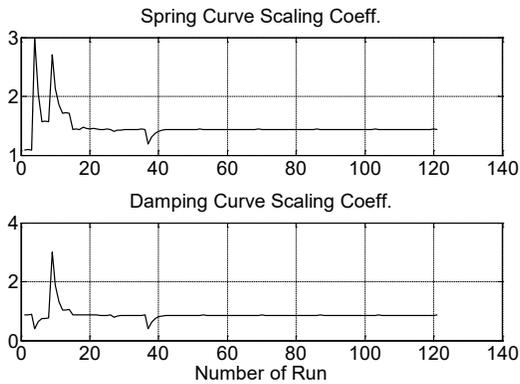
Fig. 32 Progress of spring and damper characteristic scaling coefficients

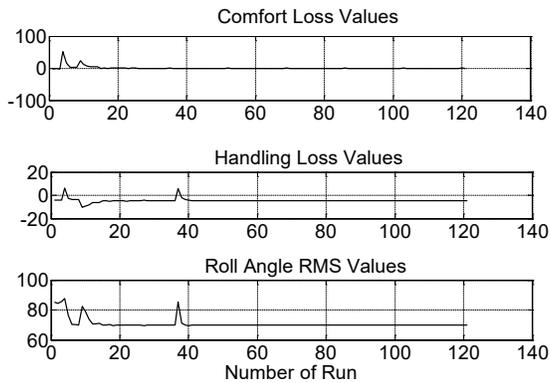
Fig. 33 Components of the objective function

As a result of optimisation, spring characteristic scaling factor tends to increase, while damper characteristic scaling factor shows a slight decrease as compared to Case 2 results. After the process, it is seen that the comfort term in the objective function worsened 9.95 % and the handling term improved 2.19 %. The improvement of the roll angle can be observed from Figure 34.

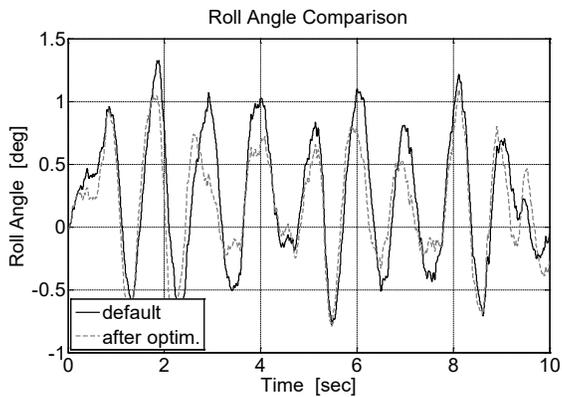

Fig. 34 Roll angle comparison on a rough road

**Handling Tests Using Carmaker Software**

Finally, the handling performances of the optimized suspension parameters are confirmed by employing the high fidelity Carmaker model of the considered vehicle (see Fig. 35). A standard double lane change maneuver is utilized to observe the body roll improvement. The simulation results (see Figs. 36 and 37) show that the nonlinear optimization has improved handling performance slightly.

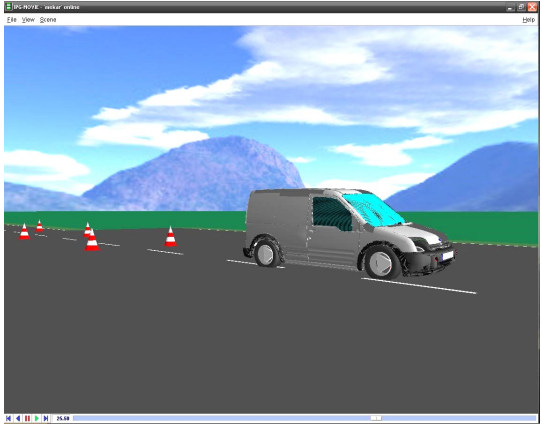

Fig. 35 Snapshot of Carmaker animation

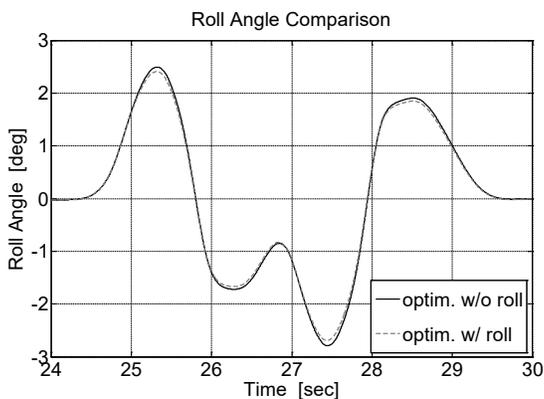

Fig. 36 Roll angle comparison double lane change

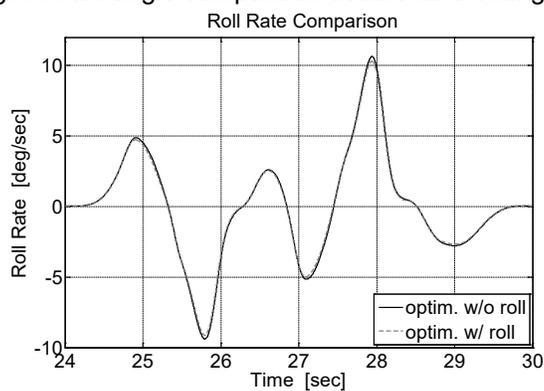

Fig. 37 Roll rate comparison double lane change

As the last investigation, the suspension units with linear and nonlinear components are compared in a double lane change maneuver using Carmaker full car model of the considered vehicle. The characteristic coefficients of the springs and dampers are used as the optimized values obtained by the linear optimisation cases while nonlinear characteristics of the components are determined by the nonlinear optimisation cases. Figures 38 and 39 show the body roll angle and roll rate comparison of the linear and nonlinear cases, respectively. As seen in the figures, nonlinear spring and damper characteristics provide better handling as compared to their linear counterparts.

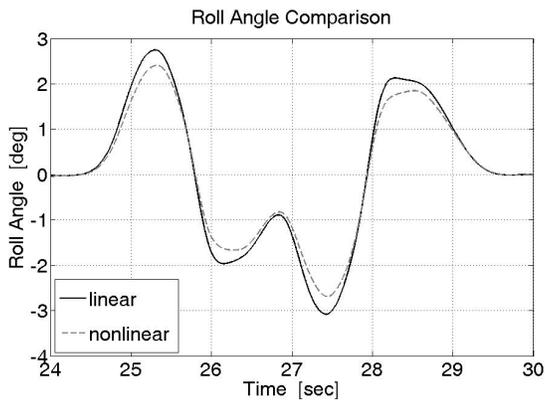
Fig. 38 Roll angle of the body during the maneuver

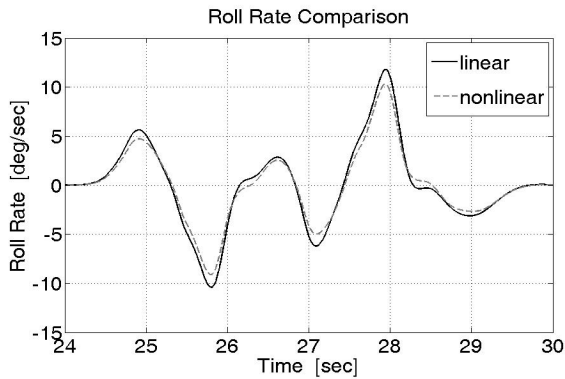
Fig. 39 Roll rate of the body during the maneuver

**CONCLUSIONS AND FUTURE WORK**

Suspension systems are a very important part of vehicle ride comfort and handling and their optimization as presented here will have effects in a large number of automotive control applications [21-91]. Their interaction and integration with suspension optimization and control would be interesting future areas of research.

Light commercial vehicle front and rear suspension units incorporating both linear or nonlinear spring and dampers were optimized to improve vehicle ride and handling. The nonlinear equations of motions of a quarter car and half car representing the front and rear suspension models were presented and simulated in the Matlab/Simulink environment.

Several aspects of performance criteria were considered for ride comfort and handling such as RMS values of weighed body acceleration, tire accelerations, the range of tire forces, RMS values of body roll angle, and deviations from a desired response etc. For each aspect of performance, time-domain performance measures were evaluated after the optimisation run.

A simple optimisation methodology of nonlinear suspension unit was presented, incorporating typical data provided by car manufacturers for the initial characteristics. The methodology was based on keeping the shape of the damper and spring properties and curve fitting a proper function to these data and then scaling it throughout the optimisation process.

Finally, the advantage of the nonlinear optimized suspension unit compared to a linear optimized suspension unit was demonstrated using a a double lane change maneuver with a high fidelity full vehicle model in Carmaker.

The more complex and more innovative optimisation problem is the shape optimisation of the nonlinear function representing the stiffness or damper characteristics. However, this shape optimisation demands a more analytical approach and it also has the following difficulties:

1. The contribution of each term of the nonlinear function could change the shape significantly and sometimes unpredictably. A global optimum points might not be achieved.

2. Springs or dampers having an unusual functional profile cannot be realized and manufactured.

In order to generalize the nonlinear suspension unit optimisation problem; an interactive MATLAB toolbox is being constructed and will be reported in the future.

**ACKNOWLEDGMENTS**

The authors thank Yıldıray Koray of İstanbul Technical University for his help in the road profile used. The authors thank Mustafa Sinal and Selçuk Kervancıoğlu of Ford Otosan for helpful discussions on suspension design. The authors also thank Ford Otosan and EU FP6 project AUTOCOM INCO-16426 for their support.

## ABBREVIATIONS

D.O.F. : Degree of freedom
$I_i$ : Performance index, i=1,2,3…
$I$ : Total index
$RMS$ : Root mean square

## LIST OF SYMBOLS

### Quarter car model

$m_s$ : Sprung mass, kg
$m_u$ : Unsprung mass, kg
$z_s$ : Sprung mass vertical displacement, m
$z_u$ : Unsprung mass vertical displacement, m
$z_r$ : Road irregularity, m
$b_s$ : Linear damper coefficient
$b_u$ : Linear tire damping coefficient
$k_s$ : Linear spring stiffness
$k_u$ : Linear tire stiffness
$B_F$ : Nonlinear damper characteristic function

### Half car model

$m_s$ : Sprung mass, kg
$m_u$ : Unsprung mass, kg
$z_s$ : Sprung mass vertical displacement, m
$z_u$ : Unsprung mass vertical displacement, m
$z_{RL}, z_{RR}$ : Left and right road irregularities, m
$b_{sL}, b_{sR}$ : Left and right linear damper coefficients, Ns/m
$k_{sL}, k_{sR}$ : Left and right linear spring stiffness, N/m
$k_{uL}, k_{uR}$ : Left and right linear tire stiffness, N/m
$B_{FL}, B_{FR}$ : Left and right nonlinear damper characteristic functions, Ns/m
$K_{FL}, K_{FR}$ : Left and right nonlinear damper characteristic functions, N/m
$L$ : Track width, m
$\phi_s$ : Sprung mass roll angle, deg/sec
$\phi_u$ : Unsprung mass roll angle, deg/sec

$I_{xx}$ : Sprung mass moment of inertia about x-axis

$I_{uxx}$ : Unsprung mass moment of inertia about x-axis

**Optimisation Process**

$w_i$ : Weighing, i=1,2,3…

$\Delta F_{tire}$ : Tire force change, N

$\Delta F_{tireopt}$ : Optimized value of tire force change, N

$\ddot{Z}_s$ : RMS of weighed sprung mass acceleration

$\ddot{Z}_{sopt}$ : Optimized value of the RMS of weighed sprung mass acceleration

$\Phi_s$ : RMS of sprung mass roll angle

$C_{lost}$ : Comfort loss

$H_{lost}$ : Handling loss